\documentclass[journal=nalefd,manuscript=letter]{achemso}
\setkeys{acs}{maxauthors=10}
\setkeys{acs}{etalmode=truncate}
\usepackage[version=3]{mhchem} % Formula subscripts using \ce{}
\usepackage{color}

\author{Emma R. Schmidgall}
\email{eschmid@uw.edu}
\affiliation[University of Washington Phys]
{Department of Physics, University of Washington, Seattle WA 98195}
\altaffiliation{E. R. S. and S. C. contributed equally to this work.}
\author{Srivatsa Chakravarthi}
\email{srivatsa@uw.edu}
\affiliation[UW EE]{Department of Electrical Engineering, University of Washington, Seattle WA 98195}
\altaffiliation{E. R. S. and S. C. contributed equally to this work.}
\author{Michael Gould}
\affiliation[University of Washington Phys]
{Department of Physics, University of Washington, Seattle WA 98195}
\author{Ian R. Christen}
\affiliation[University of Washington Phys]
{Department of Physics, University of Washington, Seattle WA 98195}
\author{Karine Hestroffer}
\affiliation[Humboldt]{Department of Physics, Humboldt-Universitat zu Berlin, 12489 Berlin, Germany}
\author{Fariba Hatami}
\affiliation[Humboldt]{Department of Physics, Humboldt-Universitat zu Berlin, 12489 Berlin, Germany}
\author{Kai-Mei C. Fu}
\affiliation[University of Washington Phys]
{Department of Physics, University of Washington, Seattle WA 98195}
\alsoaffiliation[UW EE]
{Department of Electrical Engineering, University of Washington, Seattle WA 98195}

\title[Frequency control of single quantum emitters]
  {Frequency control of single quantum emitters in integrated photonic circuits}

\keywords{diamond; nitrogen-vacancy (NV) center; integrated photonics; Stark effect; quantum emitter}

\begin{document}

%%%%%%%%%%%%%%%%%%%%%%%%%%%%%%%%%%%%%%%%%%%%%%%%%%%%%%%%%%%%%%%%%%%%%
%% The abstract environment will automatically gobble the contents
%% if an abstract is not used by the target journal.
%%%%%%%%%%%%%%%%%%%%%%%%%%%%%%%%%%%%%%%%%%%%%%%%%%%%%%%%%%%%%%%%%%%%%
\begin{abstract}
Generating entangled graph states of qubits requires high entanglement rates, with efficient detection of multiple indistinguishable photons from separate qubits. Integrating defect-based qubits into photonic devices results in an enhanced photon collection efficiency, however, typically at the cost of a reduced defect emission energy homogeneity. Here, we demonstrate that the reduction in defect homogeneity in an integrated device can be partially offset by electric field tuning. Using photonic device-coupled implanted nitrogen vacancy (NV) centers in a GaP-on-diamond platform, we demonstrate large field-dependent tuning ranges and partial stabilization of defect emission energies. These results address some of the challenges of chip-scale entanglement generation.  
\end{abstract}

%%%%%%%%%%%%%%%%%%%%%%%%%%%%%%%%%%%%%%%%%%%%%%%%%%%%%%%%%%%%%%%%%%%%%
%% Start the main part of the manuscript here.
%%%%%%%%%%%%%%%%%%%%%%%%%%%%%%%%%%%%%%%%%%%%%%%%%%%%%%%%%%%%%%%%%%%%%

%Introduction
\section*{}
A quantum network, or entangled graph state of qubits \cite{ref:raussendorf2001aow,ref:benjamin2006bgs,ref:benjamin2009pmb, ref:santori2010nqo},  is a valuable resource for both universal quantum computation and quantum communication. Optically heralded entanglement \cite{ref:barrett2005ehf} can be utilized to generate the graph state's entanglement edges between qubit nodes. However, to date, entanglement generation rates are too low to realize multi-qubit networks \cite{ref:bernien2013heb,ref:kalb2017edb, ref:hucul2014mea,ref:delteil2016ghe,ref:schwartz2016dgc} due to  photon emission into unwanted spatial and spectral modes. The integration of crystal defect-based qubits with photonic circuits offers an opportunity to significantly enhance photon collection efficiency \cite{ref:schroder2016qnd}, albeit at the cost of degrading the defect's optical properties, such as an increase in inhomogeneous emission energies and decreased spectral stability \cite{ref:faraon2011rez,ref:fu2010cnn}. Since the entanglement protocols used for generating graph state edges require detection of multiple indistinguishable photons from separate emitters \cite{ref:barrett2005ehf,ref:cabrillo1999ces}, compensating for the static and dynamic spread in emission energies is of critical importance for scalable on-chip graph state generation. 

The dc Stark effect has been used previously to tune \cite{ref:bassett2011ets,ref:bernien2013heb} and stabilize \cite{ref:acosta2012dso} the emission energy of quantum defects in bulk diamond. Here, in a series of experiments, we demonstrate the ability to tune the emission energy of photonic device-coupled near-surface NV centers over a large tuning range. Measurements on many single waveguide-coupled NV centers highlight the variability in response to an applied bias voltage, suggesting challenges in reaching the level of control necessary for chip-scale integration. Despite this variability, we are able to apply real-time voltage feedback control to partially stabilize the emission energy of a device-coupled NV center. 
%Platform 

    \begin{figure}
    \begin{center}
    \includegraphics[width =0.5\textwidth]{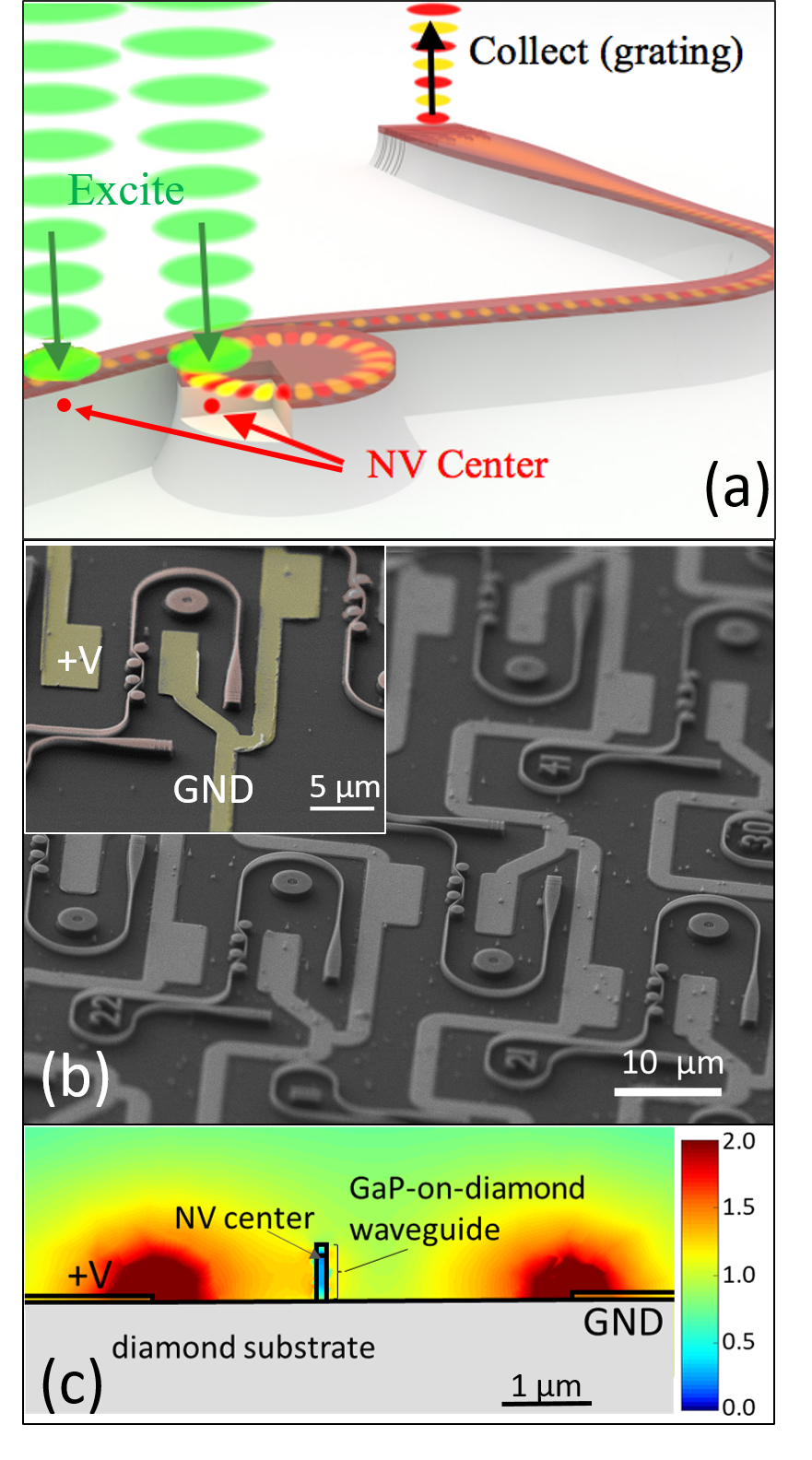}
    \caption{(a) The measurement setup. Excitation is provided by a 532~nm laser at normal incidence to either an NV center in a waveguide or a waveguide-coupled disk resonator. Collection is via the grating coupler at the end of the waveguide. (b) SEM of finished devices. (Inset) False colored SEM image (GaP = pink, Ti/Au = yellow, diamond = grey). Electrode voltages are indicated by +V and GND. (c) Cross section of the photonic devices and electrodes. The magnitude of a simulated electric field in units of 10$^7$~V/m, resulting from the application of 100~V to the +V electrode, is shown in the color scale.}
    \label{fig:device}
    \end{center}
    \end{figure}
    
%Paragraph 1: What it does
	 In a GaP-on-diamond photonics platform,\cite{ref:barclay2011hnr, ref:gould2016lsg} we are able to efficiently couple single NV centers to disk resonators, enhancing the zero-phonon line (ZPL) emission rate via the Purcell effect \cite{ref:gould2016eez}. NV centers within $\sim$ 15~nm of the diamond surface, created via implantation and annealing, couple evanescently with the GaP layer. As a result of the static dipole moment of the defect's excited state, there is variation in emission energy both between different defects, due to variation in the local environment caused by implantation and processing damage, and in the emission energy of a single defect over time due to electric field fluctuations.  However, this dipole moment also enables electric field control of the defect's emission energy \cite{ref:tamarat2006ssc,ref:bassett2011ets,ref:bernien2012tpq,ref:bernien2013heb}. We provide this control through the addition of Ti/Au electrodes to this GaP-on-diamond photonics platform. 

    In the photonic devices used in these experiments\cite{SI}, NV centers are evanescently coupled to either a 150~nm-wide GaP single-mode waveguide or a whispering gallery mode of a 1.3 $\mu$m diameter waveguide-coupled disk resonator (Figure \ref{fig:device}(a)). A grating coupler at the end of the waveguide enables collection and measurement of the NV center emission. Around each photonic device is a pair of Ti/Au electrodes with a $\sim 6~\mu$m spacing (Figure \ref{fig:device}(b))\cite{SI}. These electrodes allow application of a biasing electric field transverse to the waveguides, with a simulated field strength inside the waveguide and disk resonators of a few MV/m (Figure \ref{fig:device}(c))\cite{SI}, similar to values used in previous Stark effect tuning experiments \cite{ref:acosta2012dso}. Due to the (001) diamond growth direction, this externally applied field has components both parallel and perpendicular to the NV axis\cite{SI}. Details on device fabrication and yield can be found in Ref.[19] and in the Supplemental Information (SI)\cite{SI}.
    
    Measurements were performed between 12-14~K in a closed-cycle He cryostat. A 532~nm laser was used for optical excitation, focused onto the sample with a 0.7~NA microscope objective. Photoluminescence (PL) was collected from the grating coupler using the same objective, coupled into a grating spectrometer, and detected by a CCD camera (Figure \ref{fig:device}(a)). The input and collection optical paths were separated by a 562~nm dichoric beamsplitter. Bias voltages were applied using a computer-controlled piezocontroller in the range of 0-100~V.

%What experiments did we do
\begin{figure}
\begin{center}
\includegraphics[width=0.5\columnwidth]{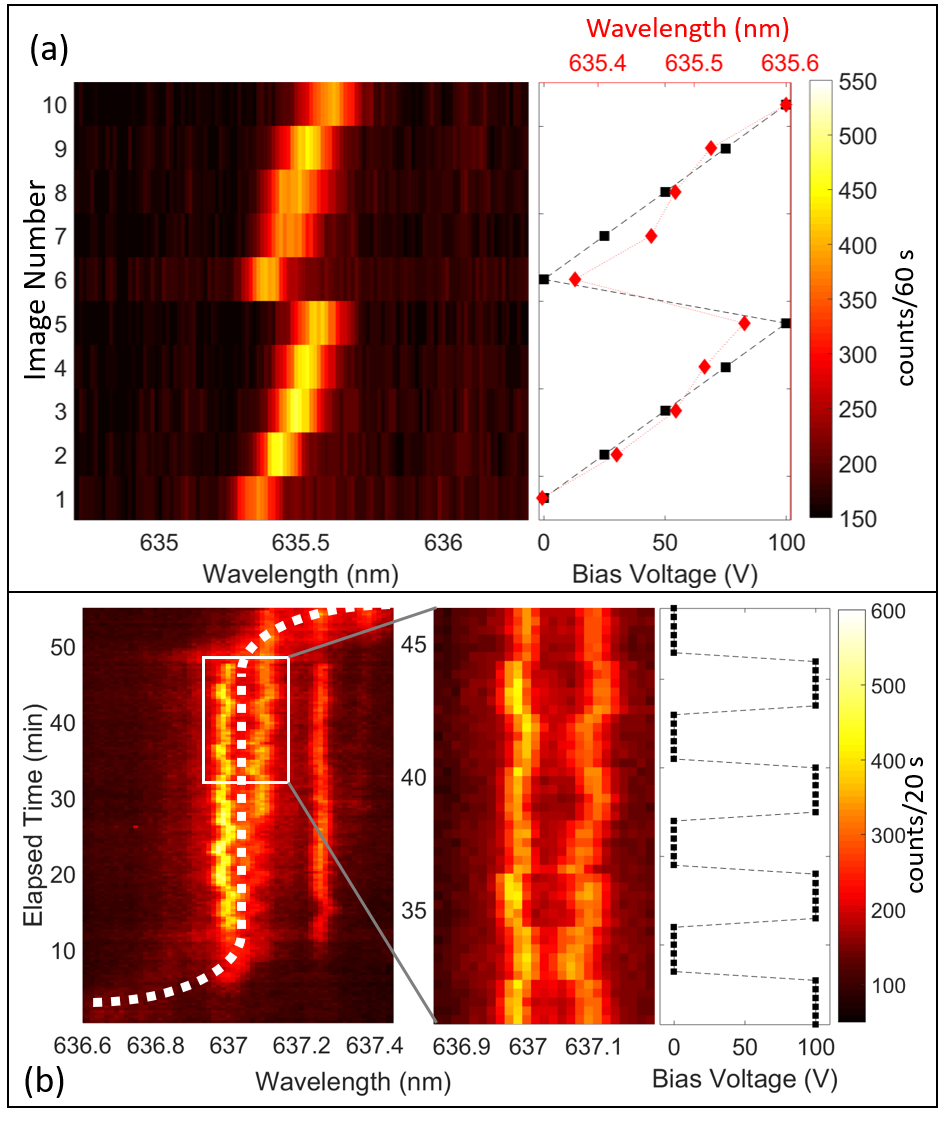}
\caption{(a) PL emission from a single waveguide-coupled ZPL showing (left) PL as a function of wavelength and (right) the applied bias voltage. The red diamonds are the center wavelength of the ZPL (upper axis). (b) Left: Xe gas tuning of a disk resonator. The cavity mode is indicated by a dashed white line. The laser excitation spot is moved slightly between the two measurements at $t\sim28$~minutes, resulting in the appearance of a second coupled NV center. At 50 minutes, the Xe gas flow is restored and the cavity tuned from resonance. The two coupled emission lines tune with the application of the bias.}
\label{fig:tuning}
\end{center}
\end{figure}

%Linear tuning of waveguide-coupled NV
We first demonstrate electric-field tuning of a waveguide-coupled NV center. Exciting the waveguide at normal incidence with collection from the relevant grating coupler (Figure \ref{fig:device}(a)), we found several coupled NV centers whose ZPL emission could be tuned with an applied bias voltage. The tuning range varied from a few GHz to a few hundred GHz. Figure \ref{fig:tuning}(a) shows an example of a nearly linear response of emission energy to applied bias voltage for a waveguide-coupled NV center with a $\sim 185$ GHz repeatable tuning range. 

We also electrically control the emission energy of a resonator-coupled NV center. We first tune the cavity mode of a waveguide-coupled disk resonator onto NV ZPL resonance via Xe gas deposition, while collecting the PL emission from the waveguide grating coupler. The Xe gas deposition results in a redshift of the resonator cavity mode. Figure \ref{fig:tuning}(b, left) shows the resulting Xe gas tuning curve for one disk resonator. Xe gas flow is halted from $t\sim15$~minutes to $t\sim45$~minutes to perform two voltage experiments and then resumed. NV centers that couple with the cavity mode are bright when in resonance with the cavity mode and not visible otherwise \cite{ref:gould2016eez}. There are several NV centers that couple to the cavity mode for this particular disk resonator. With the cavity mode tuned to resonance with two NV centers, we apply a square wave bias voltage (Figure \ref{fig:tuning}(b)), and we see the two ZPL emission lines moving in response to the applied bias voltage. These emission lines are from two separate centers since the laser spot position corresponding to maximum PL emission is different for each center. Specifically, when the laser spot position is adjusted at $t\sim28$~minutes, the second ZPL appears. This demonstrates the ability to tune NV center emission energy even while coupled to a cavity mode, combining the enhanced emission rates from the cavity coupling with tunability from the electrodes.

\begin{figure}
\begin{center}
\includegraphics[width = 0.5 \columnwidth]{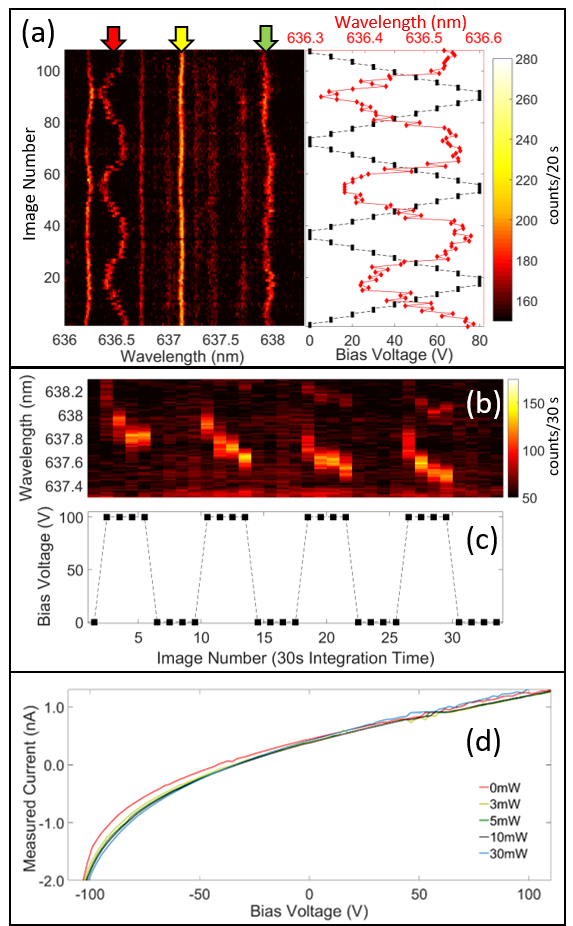}
\caption{ (a) Examples of different types of voltage-dependent behavior in an ensemble of waveguide-coupled NV centers. (Left) PL as a function of wavelength. (Right) Applied bias voltage (black squares, bottom axis) and center wavelength of the PL emission from the ZPL indicated by a red arrow in the left panel (red squares, top axis). Examples of voltage-independent ZPL emission energy (yellow arrow) and uncorrelated spectral diffusion (green arrow) are also indicated in the left panel. Color scale is in counts per 20~s. (b) ZPL emission energy as a function of time for the applied square-wave bias voltage (c). A slow transient response of the PL emission energy is observed. (d) Current-voltage characteristic for the electrodes measured for different incident green laser powers, showing the presence of a power-dependent photocurrent.}
\label{fig:strange}
\end{center}
\end{figure}

Our measurements of the tuning range of many centers show substantial variation in individual NV center voltage response. Due to the implantation density used to create the NV centers \cite{SI}, several centers can be excited in a single laser spot ($d_{laser}\sim 0.8~\mu$m). Figure \ref{fig:strange}(a) shows the bias voltage response of such an ensemble in the same waveguide. One of the NV centers, indicated by a red arrow above Figure \ref{fig:strange}(a), displays a large voltage-dependent tuning range of 190 GHz, while a nearby NV ZPL has a stable emission energy under the applied bias voltage (yellow arrow) and a third displays a large spectral diffusion ($\sim$ 70 GHz) uncorrelated with the applied bias voltage (green arrow). This variation in NV center behavior has several causes. The first factor is the center's relative orientation to the electric field. Longitudinal fields, parallel to the NV symmetry axis, will shift excited state energy levels linearly, while transverse fields will result in an excited state energy splitting that grows with increasing field.\cite{ref:bassett2011ets,ref:acosta2012dso,SI} The second factor is the few-mW non-resonant excitation, which results in photoionization of nearby defects and long-lived non-equilibrium charge distributions that can either amplify or screen the external electric field,\cite{ref:bassett2011ets} changing the effective Stark shift.\cite{ref:bassett2011ets,ref:bernien2012tpq}  The local electric field is the combination of this local photoinduced field and the externally applied field. Previous work on Stark tuning of grown-in NV centers has shown similar variation in tuning ranges and voltage responses.\cite{ref:tamarat2006ssc,ref:bassett2011ets,ref:acosta2012dso}  This variability in behavior even over small spatial scales presents a challenge to achieving the level of control required for chip-scale entanglement generation. 

In addition to spatial variability, we often observe temporal variability in the bias voltage response of these NV centers. Figure \ref{fig:strange}(b) shows a slow transient response of the ZPL emission energy from a single waveguide-coupled NV center to the applied bias voltage (Figure \ref{fig:strange}(c)). Current-voltage characteristic measurements of the electrodes, measured with a source measure unit, demonstrate the generation of a laser power-dependent photocurrent of a few pA per mW (Figure \ref{fig:strange}(d)), which suggests that slow photoinduced charging processes are responsible for these transient responses \cite{ref:bassett2011ets}. An exponential decay with a $\sim$ 40~s time constant fits the observed response. The observed emission linewidths of these NV centers also suggest fast spectral diffusion of the ZPL energy. Consequently, charging processes in this system take place over a range of timescales from sub-second to several tens of seconds. Voltage feedback stabilization must address energy variation across this range of timescales, implying that faster stabilization measurements will provide better correction to both the effects of the fast spectral diffusion and the slower field variations.

%Stabilization of NV with voltage feedback
Chip scale entanglement generation will require addressing the spatial and temporal variations of implanted NV center energies so that multiple centers can be tuned to and maintained in resonance, despite the static and dynamic energy variation of these centers that results from implantation and processing damage. Previous NV center emission energy stabilization experiments used photoluminescence excitation (PLE) measurements of grown-in NV centers for feedback control of the bias voltage \cite{ref:acosta2012dso}, but the large dynamic energy variation of implanted NV centers annealed at 800$^{\circ}$C makes PLE measurements difficult \cite{ref:fu2010cnn}. To obtain the necessary spectral resolution to demonstrate temporal stabilization, we thus utilized an Echelle spectrometer with 1.3~GHz resolution~\cite{ref:acosta2012dso}. 

\begin{figure}
\begin{center}
\includegraphics[width = 0.7\columnwidth]{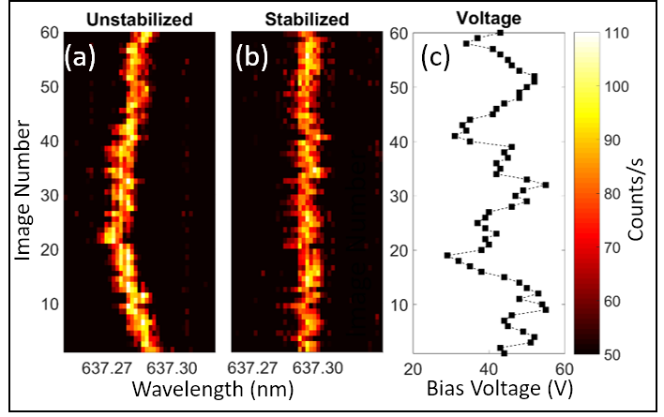}
\caption{(a) PL from an unstabilized ZPL at a constant 45~V bias voltage. The maximum spectral diffusion is 7.2~GHz and the average absolute difference from the center wavelength is 2.7~GHz. (b) Stabilized ZPL PL from a waveguide-coupled NV center. The maximum spectral diffusion is 3.9~GHz and the average absolute difference from the center wavelength is 1.2~GHz. (c) Bias voltage applied during active Stark effect stabilization of the ZPL in (b). NV PL is collected at the excitation spot due to the temporal drift of the cavity resonance over the time required for these measurements.}
\label{fig:stab}
\end{center}
\end{figure}

We identified a spectrally-diffusing NV center with a Stark tunability of $\sim$ 300~MHz/V over a range of 100~V. We measured the spectral diffusion of this NV center under a constant bias voltage of 45~V for 50 minutes (Figure \ref{fig:stab}(a)), using a 40~s integration time for the CCD to ensure an adequate signal to noise ratio. The measured linewidth was 4.5$\pm 1.2$~GHz, a result of spectral diffusion of the ZPL emission during the integration time of the CCD camera. In the second measurement, the bias voltage was actively adjusted  to stabilize the ZPL emission energy. Figure \ref{fig:stab}(c) shows the applied bias voltage as a function of time. After each 40s spectrum, the ZPL emission energy was determined and a correction to the bias voltage applied based on a PID algorithm\cite{SI}. Figure \ref{fig:stab}(b) shows the reduced ZPL spectral diffusion under active stabilization with linewidth of 5.2$\pm 1.2$ ~GHz.

%Discussion of fabrication yields
Finally, chip-scale entanglement generation in this platform will require high yields from the fabrication process. In this experiment, 12 waveguides and 21 resonators were studied. All waveguides and 8 of the resonators exhibited coupled NV centers. Of the 8 resonator-coupled centers, two could be voltage-tuned more than 20~GHz with the remainder tuning $<$ 6~GHz. This yield demonstrates the need for both NV center-device alignment, e.g. via patterned implantation \cite{ref:toyli2010csn}, and full Stark control (2 or 3 axis)\cite{ref:bassett2011ets,ref:acosta2012dso} for chip-based entanglement. 

We have demonstrated electric field control of the emission energy of photonic device-integrated implanted NV centers and feedback stabilization of this emission energy. Combined with the enhanced emission rates and collection efficiencies possible in photonic devices, these results are a necessary component of chip-scale entanglement generation. The performance of these devices can be improved by feasible steps. First, implantation and high-temperature vacuum annealing recipe development\cite{ref:chu2014cot} can result in near-surface NV centers with narrower linewidths and smaller spectral diffusion. This will enable the use of PLE measurements for the requisite fast feedback stabilization and better enable the generation of indistinguishable photons from separate NV centers. Second, future device designs can incorporate both patterned implantation and multiaxis Stark control to improve the yield of tunable NV centers. With these improvements, it should be possible to perform on-chip generation of indistinguishable photons from multiple NV centers with the collection rates necessary for scalable entangled state generation.

%%%%%%%%%%%%%%%%%%%%%%%%%%%%%%%%%%%%%%%%%%%%%%%%%%%%%%%%%%%%%%%%%%%%%
%% The "Acknowledgement" section can be given in all manuscript
%% classes.  This should be given within the "acknowledgement"
%% environment, which will make the correct section or running title.
%%%%%%%%%%%%%%%%%%%%%%%%%%%%%%%%%%%%%%%%%%%%%%%%%%%%%%%%%%%%%%%%%%%%%
\begin{acknowledgement}
This material is based on work supported by the National Science Foundation (Grant No. 1506473) and the Defense Advanced Research Projects Agency (Award No. W31P4Q-15-1-0010). We would like to acknowledge V. Acosta and C.Santori for useful discussions, N. Shane Patrick for e-beam lithography support, and HP Labs for donation of the Echelle spectrometer. Devices were fabricated at the Washington Nanofabrication Facility, a part of the National Nanotechnology Coordinated Infrastructure network. E. R. Schmidgall was supported by an appointment to the Intelligence Community Postdoctoral Research Fellowship Program, administered by Oak Ridge Institute for Science and Education through an interagency agreement between the U.S. Department of Energy and the Office of the Director of National Intelligence. 
\end{acknowledgement}

%%%%%%%%%%%%%%%%%%%%%%%%%%%%%%%%%%%%%%%%%%%%%%%%%%%%%%%%%%%%%%%%%%%%%
%% The same is true for Supporting Information, which should use the
%% suppinfo environment.
%%%%%%%%%%%%%%%%%%%%%%%%%%%%%%%%%%%%%%%%%%%%%%%%%%%%%%%%%%%%%%%%%%%%%
\begin{suppinfo}

Additional information on fabrication procedures, NV center properties, the stabilization algorithm, and additional figures. 

\end{suppinfo}

%%%%%%%%%%%%%%%%%%%%%%%%%%%%%%%%%%%%%%%%%%%%%%%%%%%%%%%%%%%%%%%%%%%%%
%% The appropriate \bibliography command should be placed here.
%% Notice that the class file automatically sets \bibliographystyle
%% and also names the section correctly.
%%%%%%%%%%%%%%%%%%%%%%%%%%%%%%%%%%%%%%%%%%%%%%%%%%%%%%%%%%%%%%%%%%%%%
\bibliography{fu_lab_bib}

\end{document}

\section{Electrostatic Modeling}
\begin{figure}
\begin{center}
\includegraphics[width=0.9\columnwidth]{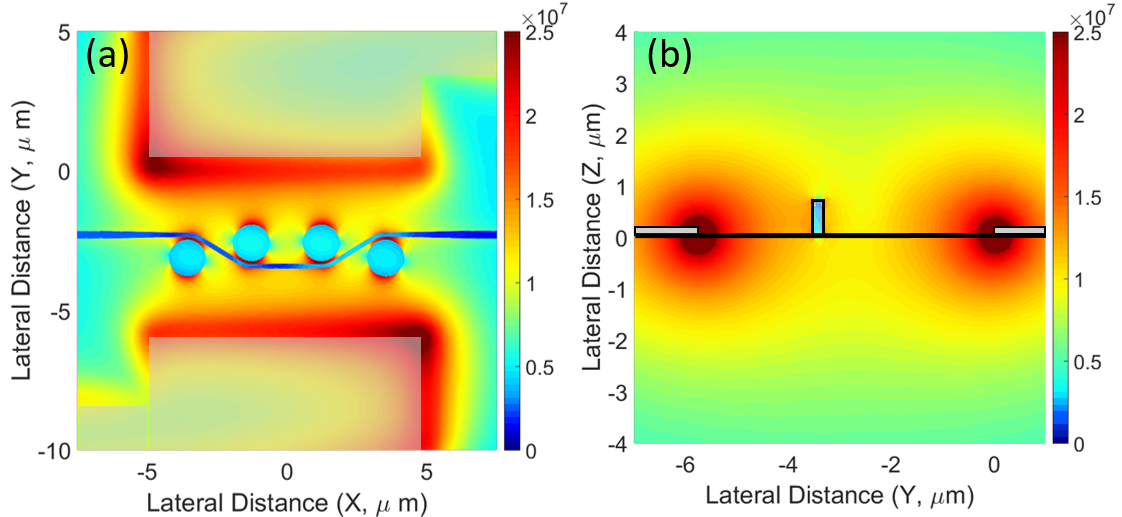}
\caption{Electric field magnitude (V/m) from the integrated electrodes for V=100~V applied to one electrode with the second electrode grounded. (a) XY (top-down) view at a height of 600~nm above the diamond substrate, corresponding to the GaP-diamond interface in the photonic devices. The semitransparent rectangles indicate the position of the electrodes on the diamond substrate. (b) YZ (side) view.}
\label{fig:sup_field}
\end{center}
\end{figure}

We modeled the electric field produced by the waveguide-transverse electrode structure in Figure 1 of the main text using COMSOL Multiphysics. Since the bottom of each electrode is composed of 7~nm Ti, we follow Acosta {\it et al.}\cite{ref:acosta2012dso} in modeling the electrodes as being composed completely of Ti. The modeled structure is incorporated directly from the GDSII file used in the electron beam lithography writes. We use a relative permittivity for the diamond substrate of $\epsilon_{r,d}=5.1$ and for the GaP $\epsilon_{r,GaP}=11.4$. We simulate 100~V applied to one electrode with the second electrode grounded, the maximum bias voltage applied in our experiments. This results in a maximum applied field inside the disk resonators of $\sim$ 5 MV/m and inside the waveguide of $\sim$ 3 MV/m at the height of the GaP/diamond interface. Figure \ref{fig:sup_field} shows the simulated field profile. It is important to note that, due to photoinduced fields especially under non-resonant excitation,\cite{ref:bassett2011ets,ref:acosta2012dso} these simulated fields are not the final electric fields inside the photonic devices. 

\section{Fabrication}
\begin{figure}
\begin{center}
    \includegraphics[width=0.8\columnwidth]{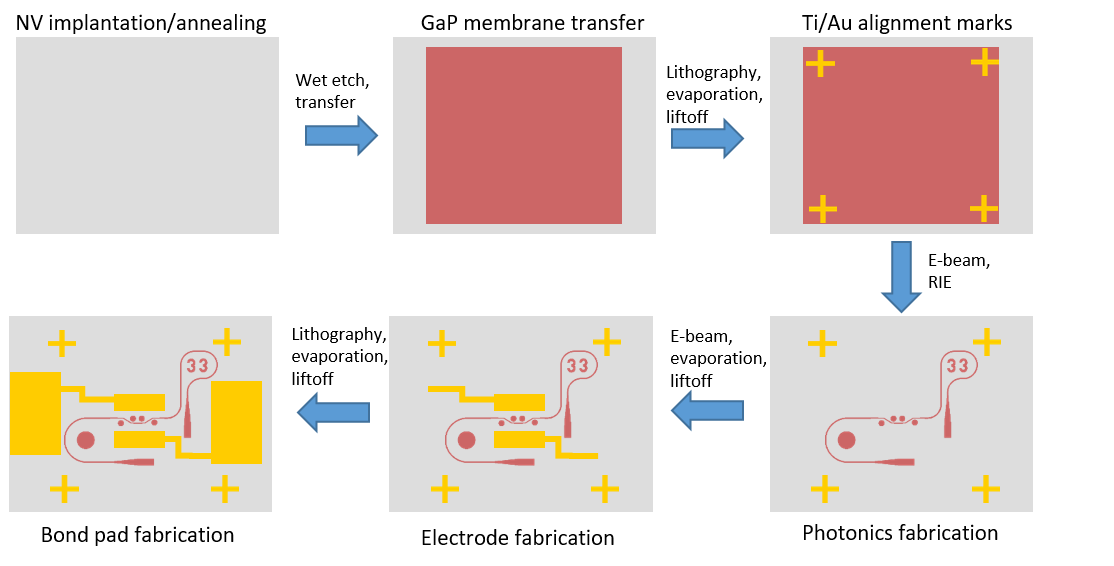}
    \caption{The fabrication process.}
    \label{fig:sup_process}
    \end{center}
    \end{figure}

Near-surface NV centers were created in the single-crystal electronic grade diamond sample (ElementSix) by N${^+}$ ion implantation (10 keV, 3$\times 10^{10}$ cm$^{-2}$, CuttingEdge Ions), followed by a two-step anneal. A 2-hour, 800$^{\circ}$C annealing step was performed under a 5\%/95\% H$_{2}$/Ar forming gas atmosphere in order to allow diffusion of vacancies to form NV centers. A subsequent 24-hour $460^{\circ}$C anneal was performed in air in order to oxygen-terminate the surface and improve stability of the negatively charged state of the near-surface NV centers \cite{ref:fu2010cnn}. A 125~nm thick GaP membrane was transferred to the diamond via epitaxial liftoff in hydrofluoric acid and van der Waals bonding \cite{ref:yablonovitch1990vdw}. Using negative-tone optical lithography (NR9-1000 resist, AD10:DI 3:1 developer), a set of 5~nm/50~nm Ti/Au alignment marks was fabricated on the GaP-on-diamond chip via evaporation and liftoff in acetone. These alignment marks were used for ensuring overlay between the electron beam lithography and reactive ion etching step that defined the photonic devices and the electron beam lithography and evaporation/liftoff step that defined the integrated Ti/Au electrodes and wires around the photonic devices. 
	Following definition of the alignment marks, hydrogen silsesquioxane (HSQ) was spin-coated to be used as the electron beam lithography resist. Photonic devices were patterned by electron-beam lithography (JEOL 6300). Two reactive-ion etch (RIE) steps were then used to etch the devices. The first RIE step (3.0 mTorr, 1.0/6.0/3.0 sccm Cl$_{2}$/Ar/N$_2$, 235~V dc bias) was used to etch the GaP layer and the second (25.0 mTorr, 20 sccm O$_{2}$, 65~V dc bias) was used to etch the diamond. The diamond etch depth was $\sim$ 600~nm. 
    For the definition of metal electrodes, PMMA A8 495 was used as a positive resist for electron beam lithography. After development in 1:1 MIBK/IPA, 7~nm/70~nm Ti/Au was deposited via evaporation and liftoff in acetone. 
    Thick 70~nm/700~nm Ti/Au bond pads for wirebonding were defined by negative-tone optical lithography (NR9-1000 resist, AD10:DI 3:1 developer), evaporation, and liftoff. The sample was then indium mounted on the sample holder, and ball/wedge ultrasonic wirebonding with Au wire was used to connect sample electrodes to the sample holder electroless nickel immersion gold PCB connections. Figure \ref{fig:sup_process} shows an overview of the fabrication process. 
    
    \section{Fabrication Yields}
    Of the 32 waveguides and 128 disk resonators fabricated on this sample, 12 waveguides had working electrodes such that a voltage-dependent response was observed from waveguide-coupled NV centers (38\%). The remaining devices had either broken waveguides (10 devices, 31\%), broken electrode leads (7 devices, 22\%), or both (3 devices, 9\%). In the intact waveguides, every waveguide had coupled NV centers whose emission could be observed via the grating coupler. Of the 128 disk resonators, 21 were located on intact waveguides with working electrodes and had cavity modes within Xe gas tuning range ($\sim$ 2nm) of the ZPL emission wavelength (16 \% of the disk resonators). Xe gas tuning resulted in resonator-coupled NV centers in 8 of these 21 resonators. Thus, the overall yield for resonator-coupled NV centers with working electrodes is 6 \%. The photonic device yields are comparable with yields reported in previous GaP-on-diamond device fabrication attempts \cite{ref:gould2016eez,ref:gould2016lsg}. The addition of electrodes lowers the overall yield for the disk resonators, but increasing the number of disk resonators per waveguide from the 4 used here to 6 or more should offset this effect. 
    
\section{Properties of the Device-Coupled NV Centers}

\begin{figure}
\begin{center}
\includegraphics[width =\columnwidth]{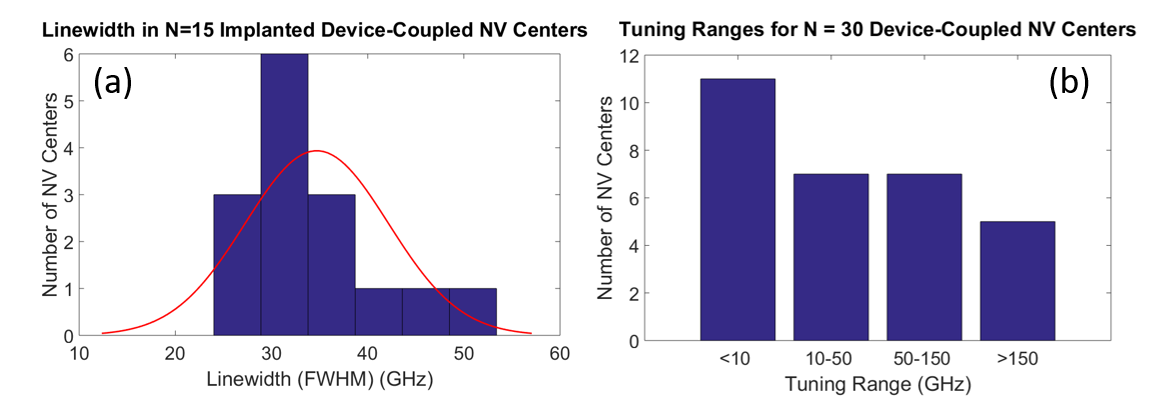}
\caption{(a) Linewidth of $N=15$ device-coupled implanted NV centers (blue bars). Red line shows a normal distribution of mean 34.7~GHz, standard deviation 7.4~GHz. (b) Tuning range of $N=30$ device-coupled implanted NV centers.}
\label{fig:NVstats}
\end{center}
\end{figure}

In these devices, the average linewidth for a device NV center as measured in our standard grating spectrometer is 34.7 GHz with a standard deviation of 7.4 GHz (Figure \ref{fig:NVstats}(a)). This is based on fitting the observed linewidths of 15 waveguide-coupled NV centers. Using our high-resolution Echelle spectrometer we find a cavity NV linewidth of 6.3 $\pm$ 3.3~GHz from a set of 14 centers. We note that due to the higher resolution of this system, we are biased toward detecting NV centers with narrower linewidths. For comparison purposes, the standard spectrometer-limited and Echelle spectrometer-limited linewidths of a single deep as-grown NV center in this same sample is 16 GHz and 1.3 GHz (1 pixel), respectively. Thus the linewidths of the device-incorporated shallow implanted NV centers are overall larger than those of growth-incorporated deep NV centers. 

We observe a broad distribution of tuning ranges (Figure \ref{fig:NVstats}(b)) within the ensemble of investigated NV centers. In a set of 30 of the observed device-coupled implanted NV centers, no tuning was visible for 11 of the centers (37\%) though this does not preclude tuning that we cannot observe with our spectrometer resolution. In 7 of the centers (23\%), tuning was observed but the range was less than 50 GHz. In 5 of the NV centers (17\%), the tuning range was between 50 and 100 GHz. In 7 of the centers (23\%), the tuning range was larger than 100 GHz. In all centers where tuning was observed, hysteretic behavior was also observed. 

\subsection{Tuning Ranges in the Absence of Strain}

\begin{figure}
\begin{center}
\includegraphics[width = 0.25\textwidth]{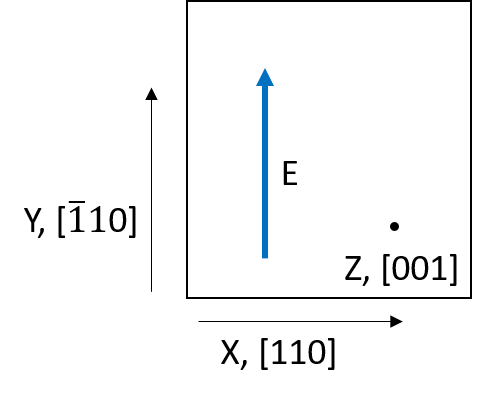}
\caption{Sample axes and field direction for estimating the tuning range.}
\label{fig:sample}
\end{center}
\end{figure}

The distribution of tuning ranges in Figure \ref{fig:NVstats}(b) is consistent with NV orientations relative to the waveguide and electrode geometry. The sample growth direction is [001] and the edges are $\langle110\rangle$ (Figure \ref{fig:sample}). 

We can define a sample $X$ axis along $[110]$ and a sample $Y$ axis along $[\bar{1}10]$. With respect to the $X$ and $Y$ axes, there are 8 possible NV center orientations: along $\pm X$ ($[111]$,$[11\bar{1}]$,$[\bar{1}\bar{1}1]$, and $[\bar{1}\bar{1}\bar{1}]$) and along $\pm Y$ (($[\bar{1}11]$,$[\bar{1}1\bar{1}]$,$[1\bar{1}1]$ and $[1\bar{1}\bar{1}]$). The photonic devices are aligned to the sample axes, so we can treat the applied electric field as being strictly along the sample $Y$ axis. Thus, for half the NV orientations, the applied field is oriented strictly perpendicular to the NV symmetry axis and results in a splitting of the $E_x$ and $E_y$ excited NV energy levels (Case 1). In the other half, the applied field has components both parallel to and perpendicular to the NV center symmetry axis, which results in both a shift and a splitting of the $E_x$ and $E_y$ energy levels (Case 2). The magnitude of these energy changes depends on the components of the NV dipole moment parallel to ($d_{||}$) and perpendicular to ($d_{\perp}$) the NV symmetry axis. The resulting energy shifts are thus: 

\begin{equation}\begin{cases}E_x = d_{\perp}F\approx 25~\text{GHz}& \text{Case 1}\\E_y = -d_{\perp}F \approx -25~\text{GHz}& \text{Case 1}\\ E_{x} = d_{||} 0.8 F+ d_{\perp} 0.6 F \approx 31~\text{GHz}& \text{Case 2}\\ E_{y} = d_{||} 0.8F - d_{\perp} 0.6 F \approx1~\text{GHz}& \text{Case 2}\end{cases}\end{equation}
where $F$ is the magnitude of the electric field and the tuning ranges are calculated using literature-reported values for $d_{||}$ and $d_{\perp}$ of 4 GHz/MV/m and 5 GHz/MV/m respectively \cite{ref:acosta2012dso} and the previously simulated field magnitude  of $F=$5 MV/m. 

These values do not take into account the fact that observed Stark tuning coefficients are $\sim $ 4 times larger under strong cw green excitation of the type used in our experiments \cite{ref:acosta2012dso,ref:bassett2011ets}, nor do they take into account photoinduced local fields. Additionally, the reported $d_{||}$, $d_{\perp}$ values are estimated to be correct only to within a factor of 2-3 \cite{ref:acosta2012dso}. The fundamental mode of the waveguide is TE polarized (parallel to the applied external field), such that only NV transitions that emit in this polarization will be observed via the grating coupler. When accounting for these factors, the tuning ranges observed in our devices are consistent with what can be expected from the applied external field. 

\subsection{Tuning Ranges in the Presence of Strain}

These implanted device-coupled NV centers are in a high strain environment, as evidenced by the wide distribution of NV center ZPL emission energies in Figure 3(a) of the main text. Consequently, the applied electric field is primarily a perturbation on the strain-induced energy variation. In the case of fixed strain, the Hamiltonian can be written as
\begin{equation}H = (\hbar \omega_0 + d_{||}F_z)\textbf{I}+\frac{1}{\sqrt{2}}\left(\begin{matrix}V_{E_x} & -V_{E_y}\\-V_{E_y} & -V_{E_x}\end{matrix}\right)\end{equation} where in this case $x$,$y$, and $z$ are the coordinate basis of the NV center including strain and $V_{E_{x,y}} =S_{E_{x,y}}-d_{\perp}F_{x,y}$ where $F_{x,y}$ is the component of the applied field along the NV center basis axes and $S_{E_{x,y}}$ are the fixed strain components \cite{ref:bassett2011ets}. The energy eigenvalues are then given by $E_{\pm}=h\nu\pm\frac{1}{2}h\delta$ where $h\nu = \hbar \omega_0+d_{||}F_z$ and $h\delta = \sqrt{2}\left(V_{E_x}^2+V_{E_y}^2\right)^{1/2}$. We can transform from the sample coordinate basis to the NV coordinate basis using the transformation matrix 
\begin{equation}\textbf{M}= \frac{1}{\sqrt{3}}\left(\begin{matrix}p_X & p_Y & -\sqrt{2}p_Z \\ -\sqrt{3}p_{Y}p_{Z}& \sqrt{3}p_{X}p_{Y}& 0\\ \sqrt{2} p_{X} & \sqrt{2}p_{Y}& p_{Z} \end{matrix}\right)\end{equation} where $p_{X,Y,Z}$ are the projections of the NV center on the sample $X$,$Y$, and $Z$ axes (eg, for [111], $p_X = 1$, $p_Y=0$ and $p_Z=1$)\cite{ref:bassett2011ets}. In the case of this [111] NV center for an applied field $F = F_Y$ along the sample $Y$ axis, the energy shifts are thus given by 
\begin{equation}E_{\pm}=\hbar\omega_0\pm\frac{\sqrt{2}}{2}\left[S_{E_x}^2+\left(S_{E_y}-d_{\perp}F_Y\right)^2\right]^{1/2}.\end{equation} Similarly, for a $[\bar{1}11]$ NV center ($p_X = 0, p_Y = 1, p_Z = 1$) and an applied field $F = F_Y$, the energy shifts are given by \begin{equation} E_{\pm} = \hbar \omega_0 + \frac{d_{||}\sqrt{2}}{\sqrt{3}}F_Y\pm\frac{\sqrt{2}}{2}\left[ \left(S_{E_x}-\frac{d_{\perp}F_Y}{\sqrt{3}}\right)^2+S_{E_y}^2\right]^{1/2}.\end{equation} In these cases, since the emission axes are primarily determined by the strain as opposed to the applied electric field, both components will couple to the TE mode of the waveguide. Thus, strain increases the number of observed emission lines and provides further variation to the expected tuning ranges.

\section{Feedback Stabilization Experiment}

The feedback was performed on the ZPL position as measured using the Echelle spectrometer. An initial tuning range measurement of the ZPL emission is used to determine the tuning range and bias voltage dependence per pixel ($V_{pix}$). The initial peak position for the initial bias voltage $p_0$ is input by the user. Initial fitting parameters (peak intensity, background level, and peak width) are also provided by the user at the beginning of the feedback measurement. 

\begin{figure}[h]
\begin{center}
\includegraphics[width = 0.7\columnwidth]{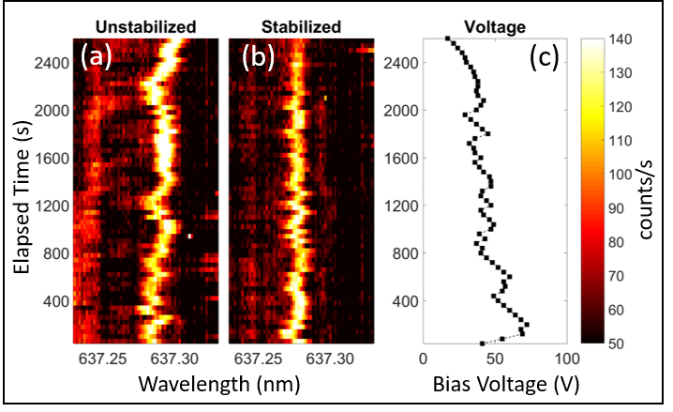}
\caption{NV frequency stabilization using proportional feedback. (a) PL from an unstabilized ZPL at a constant 45~V bias voltage. The maximum spectral diffusion is 8.3~GHz and the average absolute difference from the center wavelength is 3.2~GHz. (b) Stabilized ZPL PL from a waveguide-coupled NV center. The maximum spectral diffusion is 4.5~GHz and the average absolute difference from the center wavelength is 1.3~GHz. (c) Bias voltage applied during active Stark effect stabilization of the ZPL in (b). Feedback  is purely proportional to the error in peak position i.e. $K_p$=0.8,$~K_i$=$~K_d$=0.}
\label{fig:sup_stable}
\end{center}
\end{figure} 

\subsection{Voltage Feedback Algorithm}
The applied bias voltage is updated per spectrum based on current peak position $p$. Due to the broad linewidths, a Gaussian of the form $y = a \exp{[-(p-b)/c]^{2}}+d$ is fit to a region around the desired ZPL emission line, where $a$ is the intensity above background, $b$ the center emission energy, $c$ relates to the linewidth, and $d$ is the background. The difference between the current emission line location and the initial line location $\Delta p$ is then calculated and the bias voltage updated by a proportional integral derivative (PID) algorithm to correct for this error. \begin{equation}V_{new}=V_{prev} + (K_p~ \Delta p + K_i~I + K_d~D)~ V_{pix}\end{equation} \begin{equation}I=I + \Delta p~\Delta V \end{equation} \begin{equation}D= (\Delta p - \Delta p_{old})/\Delta V\end{equation} \begin{equation}\Delta V = |V_{prev}-V_{old}|\end{equation} where $K_p$,$~K_i$ and $K_d$ are the PID parameters determined by a semi-supervised learning algorithm. For the data shown in Fig. 4 of the main text, $K_p$=0.8,$~K_i$=$K_p$/500,$~K_d$=$K_p$/10 and $V_{pix}$=4.17 V/pixel]. $\Delta V$ is the previous voltage step and $\Delta p_{old}$ is the previous error in peak position. The PID algorithm shows higher stabilization than simple proportional feedback scheme which is shown in Fig.~\ref{fig:sup_stable}.

\bibliography{fu_lab_bib}